\documentclass{elsart}

\usepackage{amssymb}
\usepackage{amsmath}
\usepackage{latexsym}
\usepackage{graphicx}

\newcommand\I{\mathbb{I}}
\newcommand\Z{\mathbb{Z}}

\newcommand\bea{\begin{eqnarray}}
\newcommand\eea{\end{eqnarray}}
\newcommand\be{\begin{equation}}
\newcommand\ee{\end{equation}}
\newcommand\bef{\begin{figure}}
\newcommand\enf{\end{figure}}

\newcommand\lra{\longrightarrow}

\newcommand\deq{\stackrel{\mathrm{def}}{=}}

\begin{document}

\begin{frontmatter}

\title{Coherent state quantization \\ and phase operator}

\author[APC,GM]{P. L. Garc\'{\i}a de Le\'on\corauthref{cor}} \corauth[cor]{Corresponding author.}, \ead{garciadl@ccr.jussieu.fr}
\author[APC]{J. P. Gazeau} \ead{gazeau@ccr.jussieu.fr}

\address[APC]{Laboratoire Astroparticules et Cosmologie\thanksref{UMRAPC}
- Boite 7020, Universit\'e Paris 7-Denis Diderot F-75251 Paris Cedex 05, France}
\thanks[UMRAPC]{UMR 7164 (CNRS,Universit\'e Paris 7, CEA, Observatoire de Paris)}

\address[GM]{Institut Gaspard Monge (IGM-LabInfo)\thanksref{UMRGM} -
Universit\'e de Marne-la-Vall\'ee, 5 Bd. Descartes, Champs-sur-Marne,
77454 Marne-la-Vall\'ee cedex 2}
\thanks[UMRGM]{UMR 8049 (CNRS,Universit\'e de Marne-La-Vall\'ee)}

\begin{abstract}
 By using a  coherent state quantization \textit{\`a la} Klauder-Berezin,
 phase operators are constructed in  finite Hilbert subspaces of  the
 Hilbert space of Fourier series. The study of infinite dimensional
 limits of  mean values of some observables leads towards a simpler
 convergence to the canonical commutation relations.
\end{abstract}


\end{frontmatter}

\section{Introduction}

Since the first attempt by Dirac in 1927 \cite{Dir} various  definitions
of  phase operator have been proposed with more or less satisfying success
in terms of consistency \cite{SussGlo,GarrWong,PopYar,Bar,BLM}. A natural requirement is that phase operator and
number operators form a conjugate Heisenberg pair obeying the canonical
commutation relation
\be\label{ccr}
[\hat{N},\hat{\theta}] = i I_d,
\ee
in exact correspondence with the Poisson bracket for the classical
action angle variables.

To obtain this quantum-mechanical analog, the polar decomposition of
raising and lowering operators
\be\label{dec}
\hat a = \mathrm{exp}(i
\hat\theta)\hat N^{1/2},\;\;\; \hat a^{\dag} = \hat N^{1/2}\mathrm{exp}(-i
\hat \theta),
\ee was originally proposed by Dirac, with the
corresponding uncertainty relation
\be
\Delta\hat
\theta\,\Delta\hat {N}\ge \frac{1}{2}.
\ee
But the relation between operators (\ref{ccr}) is misleading. The construction of a unitary operator is a delicate
procedure and there are three main problems in it. First
we have that for a well-defined number state the uncertainty of the
phase would be greater than $2\pi$. This inconvenience,  also
present in the quantization  of the pair angular momentum-angle,
adds to the well-known contradiction lying in the matrix elements of
the commutator
\be\label{cont}
-i \delta_{n n'}=
\langle n'|[\hat  N,\hat \theta ]|n\rangle =
(n-n') \langle  n' | \hat \theta | n \rangle .
\ee
In the angular momentum
case, this contradiction is avoided to a certain extent by introducing a proper periodical variable
$\hat  \Phi(\phi)$ \cite{CarrNie}. If $\hat  \Phi$ is just a sawtooth
function, the discontinuities give a commutation relation
\be\label{deltp}
[\hat  L_z,\hat  \Phi ]= -i\{1-
2\pi\sum_{n=-\infty}^{\infty}\delta(\phi-(2n+1)\pi)\}.
\ee
The singularities in (\ref{deltp}) can be excluded, as proposed by Louisel \cite{Loui},
taking sine and cosine functions of $\phi$ to recover a valid
uncertainty relation. But the problem reveals to be harder in
number-phase case because, as showed by Susskind and Glogower
(1964)\cite{SussGlo}, the decomposition (\ref{dec}) itself leads to
the definition of non unitary operators:
\be
\mathrm{exp}(-i
\hat \theta) = \sum_{n = 0}^{\infty}|n \rangle \langle  n + 1 | \, \{ + |\psi \rangle
\langle  0 | \}, \ \mbox{and h.c.},
\ee
and this non-unitarity explains the inconsistency revealed in (\ref{cont}). To overcome this handicap, a
different polar decomposition was suggested in \cite{SussGlo}
\be
\hat a = (\hat N+1)^{\frac{1}{2}}\hat E_{-}, \;\;\; \hat a^{\dag} = (\hat N+1)^{\frac{1}{2}}\hat E_{+},
\ee
where the operators $E_{\pm}$ are still non unitary because of
their action on the extreme state of the semi-bounded number basis
\cite{CarrNie}. Nevertheless the addition of the restriction
\be\label{lowdest}
\hat  E_{-}|0\rangle = 0,
\ee
permits to define hermitian operators
\bea
\hat  C &=& \frac{1}{2}(\hat  E_{-}+ \hat  E_{+})= \hat  C^{\dag}\nonumber\\
\hat  S &=& \frac{1}{2\pi}(\hat  E_{-} - \hat  E_{+})= \hat  S^{\dag}.
\eea
These operators are named ``cosine" and ``sine" because they reproduce the same algebraic structure as
the projections of the classical state in the phase space
of the oscillator problem.

Searching for a hermitian phase operator $\hat  \theta$ which would avoid
constraints like (\ref{lowdest}) and fit (\ref{ccr}) in the classical limit,
Popov and Yarunin \cite{PopYar} and later Pegg and Barnett \cite{Bar} used an orthonormal set of
eigenstates of $\hat  \theta$ defined on the number state basis as
\be\label{phstpb}
 | \theta_m \rangle = \frac{1}{\sqrt N} \sum_{n = 0}^{N-1} e^{in \theta_m} | n \rangle.
\ee
where, for a given finite $N$, these authors selected the
following equidistant subset of the angle parameter \be
 \theta_m = \theta_0 + \frac{2 \pi m}{N}, \ m = 0, 1, \dots , N-1,
\ee
with $\theta_0$ as a reference phase. Orthonormality stems from the well-known properties of the roots of the
unity as happens with the base of discrete Fourier transform
\be
\sum_{n=0}^{N-1} e^{i n (\theta_m - \theta_{m'})} = \sum_{n=0}^{N-1} e^{i 2 \pi (m - m')\frac{n}{N} } = N \delta_{m m'}.
\ee
The phase operator on $\Cset^N$ is simply constructed through the spectral resolution
\be
\hat  \theta \equiv \sum_{m=0}^{N-1}\theta_m|\theta_m\rangle\langle \theta_m|.
\ee
This construction, which  amounts to an adequate change of orthornormal
basis in $\Cset^N$, gives for the ground number state $|0\rangle$ a random phase which avoids some of
the drawbacks in previous developments. Note that taking the limit $N\lra\infty$ is
questionable within a Hilbertian framework, this process must be
understood in terms of mean values restricted to some suitable
subspace and the limit has to be taken afterwards. In \cite{Bar} the pertinence of the states
(\ref{phstpb}) is proved by the expected value of the commutator with the number operator. The problem
appears when the limit is taken since it leads to an approximate result.

More recently an interesting approach to the construction of a phase operator has been done by Busch, Lahti and their collaborators
within the frame of measurement theory \cite{Bush}\cite{BLP}\cite{BLM}. Phase observables are constructed here using the sum
over an infinite number basis from their original definition.

Here we propose a construction based on a coherent state
quantization scheme and not on the arbitrary assumption of a
discrete phase nor on an infinite dimension Hilbert space. This will produce a suitable commutation relation at
the infinite dimensional limit, still at the level of mean values.


\section{\label{sec:level1} The approach via coherent state quantization}

As was suggested in \cite{Bar} the commutation relation will
approximate better the canonical one (\ref{ccr}) if one enlarges
enough the Hilbert space of states. We show here that there is no
need to discretize the angle variable as in \cite{Bar} to recover a suitable
commutation relation. We adopt instead the Hilbert
space $L^2(S^1)$ of square integrable functions on the  circle as
the natural framework for defining an appropriate phase operator in
a finite dimensional subspace. Let us first give an outline of the
method already exposed in \cite{gahulare,ber,klau2,gapi,gmm}.

Let $X=\{x\, \mid \, x \in X \}$ be a set equipped with a measure $\mu(dx)$
and $L^2(X, \mu)$ the Hilbert space of square integrable functions $f(x)$ on  $X$:
\bea
  \Vert f \Vert^2 = & \int_{X} \vert f(x)\vert^2 \, \mu(dx) < \infty\nonumber\\
\langle f_1 | f_2 \rangle  = & \int_{X} \overline{f_1(x)} f_2(x) \, \mu(dx)\nonumber
\eea
Let us select, among elements of $L^2(X, \mu)$,  an orthonormal set
$\mathcal{S}_N = \{ \phi_n(x) \}_{n = 1}^N$, $N$ being finite or
infinite, which spans, by definition,  the separable Hilbert subspace
${\mathcal H}_N$. We demand this set to obey the following crucial condition

\be 0 < {\mathcal
 N} (x) \equiv \sum_n \vert \phi_n (x) \vert^2 <
\infty \ \mbox{almost everywhere}. \label{factor}
\ee

Then consider the  family of states $\{ | x \rangle
\}_{x\in X}$ \underline{in} $ {\mathcal H}_N$ through the following
linear superpositions:
\be
| x\rangle \equiv \frac{1}{\sqrt{{\mathcal N} (x)}} \sum_n \overline{\phi_n (x)}
| \phi_n\rangle.
\ee
This defines an injective map (which should be continuous w.r.t some minimal topology
affected to $X$ for which the latter is locally compact):
\be
\nonumber  X \ni x
\mapsto | x \rangle \in {\mathcal H}_N,
\ee
These \textit{coherent} states obey
\begin{itemize}
  \item[¥] {\bf Normalisation}
\be
\langle \, x\, | x \rangle = 1, \label{norma}
\ee
  \item[¥] {\bf Resolution of the unity in ${\mathcal H}_N$}
\be
\int_X | x\rangle \langle x  | \, {\mathcal N}(x)\,\mu(dx)=
\I_{{\mathcal H}_N},
\ee
\end{itemize}

A {\it classical} observable is  a
function $f(x)$ on $X$ having specific properties.
Its  coherent state or frame quantization consists in
associating to $f(x)$
the operator
\be
A_f := \int_X f(x) | x\rangle \langle x| \, {\mathcal N}(x)\,\mu(dx).
\label{oper}
\ee
The function $f(x) \equiv \hat {A}_f (x)$ is called upper (or
contravariant) symbol of the operator $A_f$ and is  nonunique in general.
On the other hand,
 the mean value
$\langle x| A_f | x\rangle \equiv \check{A}_f(x)$  is called lower (or covariant) symbol of
$A_f$.

\textit{
Such a quantization of the  set $X$ is
in one-to-one correspondence with the choice of the  frame
\be
\nonumber \int_X | x\rangle \langle x  | \,{\mathcal N}(x)\,\mu(dx)=
\I_{{\mathcal H}_N}.
\ee
 To a certain extent, a
quantization scheme consists in adopting a certain point of view in
dealing with $X$ (compare with Fourier or wavelet analysis in signal processing).
Here, the validity of a precise frame choice is
asserted by comparing spectral characteristics of quantum observables
$A_f$ with  data provided by specific protocole in the observation of $X$.
}

Let us now take as a set $X$ the unit circle $S^1$ provided with the measure $\mu(d\theta)=\frac{d\theta}{2 \pi}$.
The Hilbert space is $L^2(X, \mu) = L^2 (S^1,\frac{d\theta}{2 \pi}) $ and has the inner product:
\be
\langle  f | g\rangle =\int_0^{2 \pi} \overline{f(\theta)}g(\theta) \frac{d\theta}{2 \pi}.
\ee
In this space we choose  as orthonormal set the first $N$  Fourier  exponentials with negative frequencies:
\be
 \phi_n(\theta) = e^{-in\theta}, \ \mbox{with} \ \mathcal{N}(\theta)
 = \sum_{n = 0}^{N-1} \vert \phi_n(\theta)\vert^2 = N.
\ee
The phase states are now defined as the corresponding ``coherent states'':
\be\label{phst}
 |\theta ) = \frac{1}{\sqrt{N}}\sum_{n = 0}^{N-1}  e^{in\theta} |\phi_n\rangle,
\ee
where the kets $|\phi_n\rangle$ can be directly identified to the number states $|n\rangle$, and the round
bracket denotes the continuous labelling of this family. We trivially have normalization and resolution of the
unity in ${\mathcal H}_N \simeq \Cset^N:$
\be
 ( \theta | \theta ) = 1, \ \int_0^{2\pi} | \theta )( \theta |\, N\mu(d\theta) = I_N.
\ee
Unlike (\ref{phstpb}) the states (\ref{phst}) are not orthogonal but overlap as:
\be
( \theta' |\theta ) = \frac{e^{i \frac{N-1}{2}(\theta - \theta')}}{N}\,
\frac{\sin{\frac{N}{2}(\theta - \theta')}}{\sin{\frac{1}{2}(\theta - \theta')}}.
\ee
Note that for $N$ large enough these states contain all the Pegg-Barnett
phase states and besides they form a continuous family labelled by the points of the circle.
The coherent state quantization of a particular function $f(\theta)$ with respect to the continuous
set (\ref{phst}) yields the operator $A_f$ defined by:
\be\label{Af}
 f(\theta) \mapsto \int_X f(\theta) | \theta )( \theta |\, N\mu(d\theta) \deq A_f.
\ee
An analog procedure has been already used in the frame of positive operator valued measures \cite{Bush}\cite{BLP}
but spanning the phase states over an infinite orthogonal basis with the known drawback on the convergence
of the $|\phi\rangle=\sum_n e^{in\theta}|n\rangle$ series out of the Hilbert space and the related questions concerning the
operator domain.
When expressed in terms of the number states the operator (\ref{Af}) takes the form:
\be
 A_f = \sum_{n, n' = 0}^{N-1} c_{n'-n}(f) |n\rangle \langle  n'|,
\ee
where $c_{n}(f) $ are the Fourier coefficients of the function $f(\theta)$,
\be
c_{n}(f) = \int_0^{2 \pi} f(\theta) e^{- i n \theta} \frac{d\theta}{2 \pi}.
\ee
Therefore, the existence of the quantum version of $f$ is ruled by the existence of its  Fourier transform.
Note that $A_f$ will be self-adjoint only when $f(\theta)$ is real valued. In particular, a self-adjoint phase
operator of the Toeplitz matrix type, is  obtained straightforward by choosing $f(\theta)=\theta$:
\be
 \hat  A_{\theta} = - i \sum_{{n \neq n'}\atop{n,n' = 0}}^{N-1} \frac{1}{n - n'} |n\rangle \langle  n'|,
\ee
Its lower symbol or expectation value in a coherent state  is
given by:
\be
( \theta | \hat  A_{\theta} | \theta ) = \frac{i}{N}
\sum_{{n \neq n'}\atop{n,n' = 0}}^{N-1} \frac{e^{i(n-n')\theta}}{n'
- n}.
\ee
Due to the continuous nature of the set of $|\theta)$, all  operators produced by this quantization are different
of the Pegg-Barnett operators. As a matter of fact, the commutator $[\hat {N}, \hat  A_{\theta} ]$
expressed in terms of the number basis reads as:
\be
[ \hat {N}, \hat  A_{\theta} ] =  -i \sum_{{n \neq
n'}\atop{n,n' = 0 }}^{N-1} |n\rangle \langle  n'| = i I_d + (-i)\mathcal{I}_N,
\ee
and has all diagonal elements equal to $0$. Here $\mathcal{I}_N = \sum_{n,n' = 0}^{N-1}  |n\rangle \langle  n'|$ is the $N
\times N$ matrix with all entries $=1$. The spectrum of this matrix is $0$ (degenerate $N-1$ times) and $N$.
The normalized eigenvector corresponding to the eigenvalue $N$ is:
\be
| v_N \rangle = | \theta=0 ) =  \frac{1}{\sqrt{N}}\sum_{n = 0}^{N-1}   |n\rangle
\ee
Other eigenvectors span the hyperplane orthogonal to  $| v_N \rangle$. We can choose them as the orthonormal set
with $N-1$ elements:
\be
\left\lbrace  | v_n \rangle \deq  \frac{1}{\sqrt{2}}( |n+1\rangle -  |n\rangle ), \
n = 0, 1, \dots, N-2\right\rbrace.
\ee
The matrix $\mathcal{I}_N$ is just $N$ times the projector $| v_N \rangle \langle  v_N |.$ Hence the commutation rule reads as:
\be\label{ccrjp1}
[ \hat {N}, \hat
A_{\theta} ] = -i \sum_{{n \neq n'}\atop{n,n' = 0}}^{N-1} |n\rangle \langle
n'| = i\left( I_d - N | v_N \rangle \langle  v_N |\right).
\ee
A further analysis of this relation through its lower symbol provides, for the matrix $\mathcal{I}_N$, the function:
\be
( \theta | \mathcal{I}_N
| \theta ) = \frac{1}{N}\sum_{n,n' = 0}^{N-1} e^{i(n-n')\theta} =
\frac{1}{N}
\frac{\sin^2{N\frac{\theta}{2}}}{\sin^2{\frac{\theta}{2}}}.
\ee
 In the limit at large $N$ this function is the Dirac comb (a well-known
result in diffraction theory):
\be
\lim_{N \to \infty} \frac{1}{N}
\frac{\sin^2{N\frac{\theta}{2}}}{\sin^2{\frac{\theta}{2}}} = \sum_{k
\in \Z} \delta (\theta - 2 k \pi).
\ee
Recombining this with expression (\ref{ccrjp1}) allows to recover the canonical commutation rule:
\be\label{expcr}
 (  \theta | [ \hat {N}, \hat  A_{\theta} ] | \theta ) \approx_{N \to \infty} i
 - i \sum_{k \in \Z} \delta (\theta - 2 k \pi).
\ee
This expression is the expected one for any periodical variable as was seen in (\ref{deltp}).
It means that in the Heisenberg picture for temporal evolution
\be
\hbar\frac{d}{dt}\langle \hat  A_{\theta}\rangle = -i\langle [ \hat {N}, \hat  A_{\theta} ]\rangle=1-\sum_{k \in \Z} \delta (\theta - 2 k \pi)
\ee
 A Dirac commutator-Poisson bracket correspondence
can be established from here. The Poisson bracket equation of motion for the phase of the harmonic oscillator is:
\be
\frac{d\theta}{dt}=\{H,\theta\}=\omega(1-\delta (\theta - 2 k \pi)),
\ee
where $H=\frac{1}{2}(p^2+\omega^2 x^2)$ is the Hamiltonian and $\theta= \arctan(p/{\omega x})$ is the phase. The
identification $[ \hat {N}, \hat  A_{\theta} ]=i\hbar\omega\{H,\theta\}$ is straightforward and we recover a
sawtooth profile for the phase variable just as happened in (\ref{deltp}) for the angle variable.

Note that relation (\ref{expcr}) is found through the expected value over phase coherent states and not in any physical
state as in \cite{Bar}. This shows that states (\ref{phst}), as canonical coherent states, hold the closest to classical
behavior. Another main feature is that any of these states is equal-weighted over the number basis which confirms a total
indeterminacy on the eigenstates of the number operator and the opposite is also true. A number state is equal weighted
over all the family (\ref{phst}) and in particular this coincides with results in \cite{Bar}.

The creation and annihilation operators are obtained using first the quantization (\ref{Af}) with $f(\theta)=e^{\pm i\theta}$:
\be
\hat A_{e^{\pm i \theta}}=\int_0^{2\pi} e^{\pm i\theta} N | \theta )( \theta |\frac{d\theta}{2\pi},
\ee
and then including the number operator as $\hat A_{e^{i \theta}}\hat N^{\frac{1}{2}}\equiv \hat a$ in a similar way to
\cite{Bar} where the authors used instead $e^{i\hat\theta_{PB}}\hat N^{\frac{1}{2}}$.
The commutation relation between both operators is
\be
[\hat a, \hat a^{\dagger}]= 1- N|N-1\rangle\langle N-1|,
\ee
which converges to the common result only when the expectation value is taken on states where extremal state component
vanish as $n$ tends to infinity.

As the phase operator is not built from a spectral decomposition, it is clear that
$\hat A_{\theta^2}\neq \hat A_{\theta}^2$ and the link with an uncertainty relation is not straightforward as in
\cite{Bar}, instead, as is suggested in \cite{BLP}, a different definition for the variance should be used.

The phase operator constructed here has most of the advantages of the Pegg-Barnett operator but allows
more freedom within the Hilbertian framework. It is clear that a well-defined phase operator must be
parametrised by all points
in the circle in order to have a natural convergence to the commutation relation in the classical limit.
It remains also clear that as in any measure, like Pegg-Barnett's or this one through coherent sates, the
inconveniences due to the non periodicity of the phase pointed in \cite{SussGlo} are avoided from  the very
beginning in the choice of $X \equiv S^1$.
\ack
This work was supported in part by the Consejo Nacional de Ciencia y Tecnolog\'{\i}a (CONACyT). The
authors acknowledge the referees for their constructive comments.

\end{document}